\documentclass[conference]{IEEEtran}
\IEEEoverridecommandlockouts
\usepackage{caption}
\usepackage{subfigure}
\usepackage{graphics} % for pdf, bitmapped graphics files
\usepackage{epsfig} % for postscript graphics files
\usepackage{amsfonts,amssymb}
\usepackage{multirow}
\usepackage{amsmath}
\graphicspath{{figure/}}

\begin{document}

\title{\LARGE \bf
Training Enhancement of Deep Learning Models
for Massive MIMO CSI Feedback with Small Datasets \\
\thanks{Z. Liu is with the School of Computer Science (National Pilot Software Engineering School), Beijing University of Posts and Telecommunications, China (e-mail: lzyu@bupt.edu.cn).}
\thanks{Z. Ding is with the Department of Electrical and Computer Engineering, University of California at Davis, USA (e-mail: zding@ucdavis.edu).} 
}
% \author{A, B, C}

\author{\IEEEauthorblockN{Zhenyu Liu, \emph{Member, IEEE}, and Zhi Ding, \emph{Fellow, IEEE}} 
}

\maketitle

\begin{abstract}

Accurate downlink channel state information (CSI) is vital to achieving high spectrum efficiency in massive MIMO systems. Existing works on the deep learning (DL) model for CSI feedback have shown efficient compression and recovery in frequency division duplex (FDD) systems. However, practical DL networks require sizeable wireless CSI datasets during training to achieve high model accuracy. To address this labor-intensive problem, this work develops an efficient training enhancement solution of DL-based feedback architecture based on a modest dataset by exploiting the complex CSI features, and  augmenting CSI dataset based on domain knowledge. We first propose a spherical CSI feedback network, SPTM2-ISTANet+, which employs the spherical normalization framework to mitigate the effect of path loss variation. We exploit the trainable measurement matrix and residual recovery structure to improve the encoding efficiency and recovery accuracy. For limited CSI measurements,  we propose a model-driven lightweight and universal augmentation strategy based on decoupling CSI magnitude and phase information, applying the circular shift in angular-delay domain, and randomizing the CSI phase to approximate phase distribution. Test results demonstrate the efficacy and efficiency of the proposed training strategy and feedback architecture for accurate CSI feedback under limited measurements.

\end{abstract}

\begin{IEEEkeywords}
Massive MIMO, CSI feedback, data augmentation, deep learning, FDD
\end{IEEEkeywords}

% \fi
\IEEEpeerreviewmaketitle
\vspace*{-3mm}

\section{Introduction}
Modern wireless communication systems have made tremendous strides in utilizing the spatial diversity afforded by multiple-input multiple-output (MIMO) transceivers to improve radio link performance. In particular, massive MIMO systems have shown great promise for delivering high spectrum and energy efficiency for 5G 
wireless systems and beyond. The efficiency of massive MIMO downlink depends on accurate downlink CSI estimates at gNodeB (gNB) for transmission precoding. 
For massive MIMO systems, such feedback data can be substantial because the large number of antennas and wide bandwidth lead to very high CSI dimensionality. This challenge strongly motivates many
research efforts aimed at accurate downlink CSI feedback in frequency division duplex (FDD) systems.

To reduce the bandwidth required
for CSI feedback, compressed sensing (CS)-based approaches can exploit channel properties including low rank or sparsity in spatial domain \cite{cs5_3} and temporal domain \cite{cs3_3} to derive a compressed CSI representation for feedback.  However, CS approaches rely
on strong channel sparsity which may not strictly hold in some cases, and thus can limit their efficacy \cite{ref:csinet}. The popularity and versatility of deep learning (DL) have
motivated a number of recent works that explored
deep neural networks for downlink CSI feedback. By exploiting spatial and spectral correlation \cite{ref:csinet,dl_multires, guo2019convolutional,tang2021dilated, ref:SRNet}, %[5-8] ref:AnciNet, ref:dl_multires \cite{ref:csinet, guo2019convolutional,tang2021dilated, ref:SRNet}[3-6]
bi-directional correlation \cite{ref:dualphase}, and temporal correlation \cite{ref:csinet-lstm,liu2022}, DL-based CSI feedback has 
demonstrated high recovery accuracy and time efficiency, and has been regarded as a potential technology beyond 5G.
However, the training of deep neural networks (DNNs) requires a large number of channel measurements to achieve reliable model accuracy.
Collecting numerous CSI measurements can be
quite costly and time-consuming. 

The practical challenges of channel measurement acquisition motivate the development of effective data augmentation solutions to enhance the training of DL networks. 
By modifying existing data or creating newly synthetic data, data augmentation can help mitigate overfitting when training a deep learning model. Nevertheless, traditional data augmentation techniques commonly used in image processing such as geometric transformations, cropping and rotation are incompatible with CSI estimation and feedback, since they could significantly alter CSI statistics such as delay spread distribution and would be inconsistent with physical CSI characteristics. To imitate wireless CSI features,  generative adversarial network (GAN)-based channel modeling methods have been considered for the single channels \cite{ref:GAN_yang,ref:GAN_ye} and MIMO channels \cite{ref:GAN_xiao}. Ironically, GAN often requires training by a large number of CSI measurements. Moreover, the channel generator of GAN designed for massive MIMO \cite{ref:GAN_xiao} can be computationally intensive and requires billions of floating point operations (FLOPs). 
% In addition, existing works mainly focus on , which are not suitable for the massive MIMO wideband CSI feedback. 

Besides data augmentation, to improve CSI recovery accuracy based only on limited available data samples,  DNN should be capable of handling complex CSI features and variations. Generally, 
existing DL-based CSI feedback works \cite{guo2019convolutional,ref:SRNet,ref:dualphase} have demonstrated satisfactory performance for indoor CSIs but tend to
be less effective for complex outdoor CSIs. Furthermore, CSI matrices of wide bandwidth and massive antennas tend to exhibit a high
level of variation. Meanwhile, path attenuation can make the power of CSI matrices vary by several orders of magnitude. Such inherent CSI characteristics can be particularly problematic for DNN-based CSI compression and feedback system that is trained by the small dataset of measured CSI without sufficient representation. 

% To improve the performance of DLN-based wideband massive MIMO CSI feedback for small dataset case, in this paper, we build a framework which starts from strengthening the CSI feedback network to handle the data samples with complex features, and enhancing the channel data augmentation with the help of domain knowledge.

To improve the performance of wideband massive MIMO CSI feedback given limited channel measurements, we develop an efficient DL-based feedback architecture amenable to enhanced training despite a small training dataset. Our objectives are to enhance the recovery accuracy of the CSI feedback network in the complicated environment and to develop simple data augmentation with physical insights. We design an optimized CS-inspired feedback network structure to process CSI samples with complex features. We utilize physical insights and domain knowledge regarding wireless channels to devise a lightweight but efficient CSI data augmentation technique. Our contributions are summarized as follows:

% In order to improve CSI recovery accuracy and reduce  feedback payload, we develop a novel DL framework based on Markovian learning model, MarkovNet. 
% MarkovNet systematically exploits temporal channel correlation 
% characteristics to achieve a much smaller
% model size, thereby substantially reducing
% computational complexity and memory
% requirements.

% 

\begin{itemize}
    \item To improve CSI feedback accuracy and efficiency in complex environments, we propose an efficient CS-inspired CSI feedback framework, named SPTM2-ISTANet+, where a spherical feedback structure regulates the input distribution and lessens the impact of path loss. A deep unfolding-based decoding network with residual recovery structure improves CSI recovery accuracy, with which we construct a trainable measurement matrix-based encoding network to overcome shortcomings of fixed sparse transformation in CS and improve encoding efficiency. Simulation results demonstrate superior performance, e.g., normalized mean square error (NMSE) of $-24.3$dB when the compression ratio (CR) is $\frac{1}{4}$ in a commonly used outdoor scenario \cite{ref:csinet}.
    
    % \item To overcome the influence of limited available channel measurements, different from previous works, we build a novel data augmentation approach hybridGAN, which 
    % take use of the different features in the magnitude and phase of CSI matrices separately.
    
    \item  Instead of training a black box GAN to generate augmented CS samples, we develop a simple but effective model-driven augmentation strategy by exploiting physical knowledge and features of CSI matrices based on domain insight. Taking into consideration
of geographic continuity and delay property of MIMO channels,
we develop a circular shifting augmentation of CSI magnitudes in 
angular-delay domain to incorporate the circular discrete Fourier transform (DFT) property. 
We also apply uniform phase random variation in CSI coefficients to imitate channel phase changes and mitigate overfitting.
% to imitate carrier phase jitters in practical transceivers. 
% Mitigating overfitting,  construct a larger phase constraint space than the limited measurement to
Simulation results demonstrate that the proposed augmentation for training can significantly enhance CSI recovery performance than GAN, and can achieve NMSE of $-15$dB using only $100$ real samples when CR is $\frac{1}{4}$ in the outdoor scenario.

\item We analyze the effect of different modules in the proposed deep unfolding-based CSI feedback network and channel data augmentation method to optimize the DL-based CSI encoder-decoder architecture when there exists only a small set of CSI measurement data samples in practice. 
    
    % Complexity comparison shows that the proposed augmentation method only requires thousands of FLOPs to achieve the outstanding performance, which can facilitate the deployment of DLN-based CSI feedback.
    
    % of CS-based methods.
    
    % -13dB using 100 real samples
    % \item Simulation results demonstrate that TMM-ISTANet+ can outperform the existing methods, and can achieve the -22.7dB in the outdoor scenario when the compression ratio is $\frac{1}{4}$.

\end{itemize}

% \vspace*{-2mm}
\section{System Model}

Without loss of generality, we consider a massive MIMO gNB equipped with $N_b \gg 1$ antennas
to serve a number of single-antenna UEs within its cell. 
Orthogonal frequency division multiplexing (OFDM) is adopted
in downlink transmission
over $N_f$ subcarriers. 
Let $\mathbf{h}_{m} \in \mathbb{C}^{N_b\times1}$ denote the channel vector in the $m-$th subcarrier, 
$\mathbf{w}_{m} \in \mathbb{C}^{N_b\times1}$ denote transmit precoding vector, $x_{m}\in \mathbb{C}$ 
denote the transmitted data symbol, and $n_{m}\in \mathbb{C}$ denote the additive noise.
Then the received signal of the UE on the $m-$th subcarrier  is given by
\begin{equation}
	y_{m} =\mathbf{h}_{m}^H\mathbf{w}_{m}x_{m} + n_{m}, 
\label{equ1}
\end{equation}
where $(\cdot)^H$ represents the conjugate transpose.
The downlink CSI matrix in the spatial-frequency domain is denoted by $\tilde{\mathbf{H}} = \left[\mathbf{h}_{1},..., \mathbf{h}_{N_f}\right]^H \in \mathbb{C}^{N_f\times N_b}$.  

To reduce feedback overhead, we first exploit the sparsity of CSI in the delay domain.
Applying 2D DFT, CSI matrix 
$\mathbf{H}_{sf}$ in 
spatial-frequency domain can be transformed  to be $\mathbf{H}_{ad}$ in angular-delay domain using
\begin{equation}
	\mathbf{F}_d^H \mathbf{H}_{sf}  \mathbf{F}_a =  \mathbf{H}_{ad}, \label{idft3}
\end{equation}%
where $\mathbf{F}_d$ and $\mathbf{F}_a $ denote the $N_f \times N_f$ and $N_b \times N_b$ unitary DFT matrices, respectively.  
Owing to limited multipath delay spread and scatters in practical radio environment, most elements in the $N_f\times N_b$
matrix $\mathbf{H}_{ad}$ are negligibly small except for the
first $R_d$ rows \cite{ref:csinet}. Therefore, we can approximate the
channel by truncating CSI matrix to the first $R_d$ rows, and utilize $\mathbf{H}$ to denote the truncated matrix. % of $\mathbf{H}_{ad}$

We then vectorize the truncated downlink CSI matrix $\mathbf{H}$ as input to DNN for compression. Real part and imaginary part are split for easier processing. The corresponding vector is denoted as $\mathbf{x} \in \mathbb{R}^{N}$, where $N = 2 \times R_d \times N_b$. To compress the length of vector $\mathbf{x}$, the measurement matrix $\Phi \in \mathbb{R}^{M \times N}$ are used for dimension compression. Assuming the feedback procession are lossless \cite{ref:csinet,ref:Lu2020CRNet}, the low-dimension vector $\mathbf{y} \in \mathbb{R}^{M}$ received by the gNB can be defined as $\mathbf{y}=\Phi\mathbf{x}$.

After receiving the compressed vector $\mathbf{y}$, the CSI decoder at gNB can reconstruct the original vector $\mathbf{x}$ by solving the following compressive sensing  recovery problem:
\begin{equation} \label{eq:ista}
    \min _{\mathbf{x}} \frac{1}{2}\|\Phi \mathbf{x}-\mathbf{y}\|^{2}+\lambda\|\mathcal{F}(\mathbf{x})\|_{1},
\end{equation}
where  $\lambda$ is the regularization parameter, $\|\cdot\|$ denotes the $l_2-$norm. $\mathcal{F}(\cdot)$ is the sparse transform function of $\mathbf{x}$, which can utilize the wavelet, DCT, neural networks, among others. 

% In this paper, neural networks are used to optimize the parameters of sparse transform.

% To enhance the CSI feedback performance given the limited channel measurement, in this paper, we construct the few-shot learning CSI framework from the following two perspectives:
% \begin{itemize}
%     \item The CSI feedback framework should provide high-accuracy recovery even in the complex channel environment. To achieve this goal, on the one hand, a powerful decoding network is required to handle the elaborate recovery. On the other hand, the constructed CSI feedback network should has the ability to decrease the influence of channel variability.
%     \item The data augmentation should split the most significant features in MIMO channel to decrease the number of required real measurements, and cover the cases which has not been included in the measurements.
% \end{itemize}

Next, we construct a DL-based architecture to enhance CSI feedback performance from limited CSI measurement samples from the perspectives of efficient feedback and training enhancement. 
%  \vspace*{-2mm}
\section{SPTM2-ISTANet+}
%  \vspace*{-1mm}
We construct an efficient CS-inspired deep unfolding network to improve CSI feedback
efficiency or recovery performance in this section.

\begin{figure*}[thpb]
      \centering
      \includegraphics[scale=0.36]{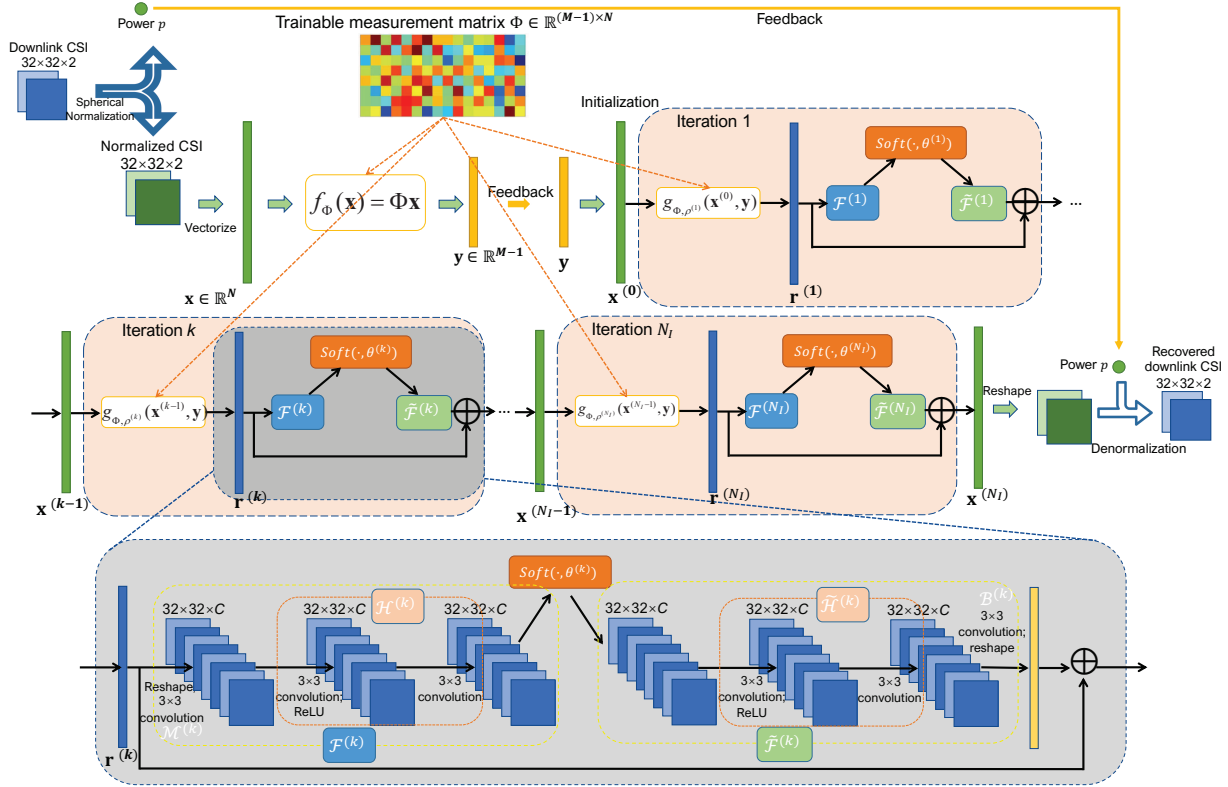}\vspace*{-1mm}
      \caption{Architecture of SPTM2-ISTANet+}
      \label{figure_architecture}
      \vspace*{-5mm}
  \end{figure*}

\vspace*{-2mm}
\subsection{Encoding Network}
 \vspace*{-1mm}

We adopt a deep unfolding-based CSI feedback network by considering physical channel features, and propose two key changes to the encoding network. 

% , as many existing DL based works mainly utilize the DL architectures successfully developed for image processing areas. 

% In this paper, to improve CSI feedback accuracy,  we adjust the deep neural network  ISTANet+ designed in the work in \cite{istanet} for image reconstruction considering the channel features, and further make two key changes in the encoding network to improve the CSI feedback accuracy evidently.

First, we construct a spherical CSI 
feedback structure in view of domain-specific characteristics of the wireless channel. 
Unlike pixel values of image data, the distribution of MIMO CSI coefficients is
substantially different. The dynamic range of CSI is always much greater because of 
radio path loss, since CSI of one user equipment (UE) may differ from CSI of another UE by orders of magnitude.
% For example, RF pathloss grows polynomially with distance 
% between gNB and UE. 
% As a result, CSI of one UE can be different from CSI of another UE by several orders of magnitude,
% depending on their respective distances to gNB. 
A naive processing can render CSI of some
UE too small, leading to large recovery errors. 
Consequently, before applying 
CS measurement matrix to lower
CSI dimension, we split the CSI matrix $\mathbf{H}_{k}$ into a power value $p_k$ and a 
spherical matrix $\check{\mathbf{H}}_{k}$, where 
$p_k = \Arrowvert\mathbf{H}_{k}\Arrowvert$ is the power
of the CSI matrix and $\check{\mathbf{H}}_{k} = \mathbf{H}_{k}/\Arrowvert\mathbf{H}_{k}\Arrowvert$
is the unit norm spherical CSI.
% Note that the elements of $\check{\mathbf{H}}_{k}$ remain strictly 
% inside or on the surface of the unit hyper-sphere.
As shown in Fig. \ref{figure_architecture}, UE first extracts and feeds back
the power $p_k$, before using the measurement matrix $\Phi \in \mathbb{R}^{(M-1) \times N}$ for dimension compression of spherically normalized downlink CSI.

Next, to deliver further performance improvement, we move away from the traditional random constructed measurement matrix and devise a data-driven trainable measurement matrix.  The goal is to better capture key CSI  features of massive MIMO for encoding, particularly if the compression degree is high such that the CR is small. Since only a matrix multiplication is used by the encoder, the computation cost at the UE is modest.

%  and smaller than \cite{ref:csinet,dl_multires, guo2019convolutional,tang2021dilated, ref:SRNet}
%, and lower than compared with \cite{ref:csinet,dl_multires, guo2019convolutional,tang2021dilated, ref:SRNet}.
% Consequently, as shown in Fig. \ref{figure_architecture}, the encoder first extracts and sends back
% the power $p$ of the downlink CSI matrices, and vectorize the rest spherically normalized downlink CSI matrices. Then, a trainable measurement matrix $\Phi \in \mathbb{R}^{(M-1) \times N}$ is utilized for dimension compression.
\vspace*{-1mm}
\subsection{Decoding Network}
\vspace*{-1mm}
Our decoding learning network of SPTM2-ISTANet+ utilizes the deep unfold structure. By
mainly adopting the settings of ISTANet+ \cite{istanet} to unfold the iterative shrinkage-thresholding algorithm (ISTA) \cite{beck2009fast}, we 
recover the CSI
by iterating between the following steps:%\cite{beck2009fast, ista_org}
\begin{equation} \label{eq:r}
    \mathbf{r}^{(k)}=\mathbf{x}^{(k-1)}-\rho \Phi^{\top}\left(\Phi\mathbf{x}^{(k-1)}-\mathbf{y}\right),
\end{equation}
\vspace*{-5mm}
\begin{equation} \label{eq:x}
    \mathbf{x}^{(k)}=\underset{\mathbf{x}}{\arg \min } \frac{1}{2}\left\|\mathbf{x}-\mathbf{r}^{(k)}\right\|^{2}+\lambda\|\mathcal{F}(\mathbf{x})\|_{1},
\end{equation}
where $k$ denotes the iteration index, $\rho$ denotes the step size. Next, Eq. (\ref{eq:r}) and Eq. (\ref{eq:x}) are respectively expanded into a deep unfolding module corresponding to the $k$ iteration, i.e. $\mathbf{r}^{ (k)}$ module and $\mathbf{x}^{(k)}$ module, to solve the recovery problem.%accurately

% SPTM2-ISTANet+ uses a combination of two convolutional layers and a Rectified Linear Unit (ReLU) $\text{ReLU}(x)=\max(0, x)$ to construct a sparse transformation $\mathcal{F}(\cdot )$, i.e., $\mathcal{F}(\mathbf{x})=\mathbf{B} \text{ReLU}(\mathbf{A} \mathbf{x})$, where $\mathbf{A}$ and $\mathbf{B}$ both use convolutional layers without bias terms to perform equivalent matrix operations.

$\mathbf{r}^{(k)}$ module corresponds to Eq. (\ref{eq:r}) and generates $\mathbf{r}^{(k)}$ from the result of the $(k-1)$-th iteration. In order to improve the flexibility of recovery network, the step size $\rho$ in the Eq. (\ref{eq:r}) may be automatically adjusted according to iteration, i.e., $\rho^{(k)}$ varies in each $\mathbf{r}^{(k)}$ module. Therefore, $\mathbf{r}^{(k)}$ module can be regarded as a function of $\mathbf{x}^{(k-1)}$ and $\mathbf{y}$, i.e.,
\begin{equation} \label{eq:rk}
    \mathbf{r}^{(k)}=g_{\Phi,\rho^{(k)}}\left(\mathbf{x}^{(k-1)}, \mathbf{y}\right)=\mathbf{x}^{(k-1)}-\rho^{(k)} \Phi^{\top}\left(\Phi\mathbf{x}^{(k-1)}-\mathbf{y}\right).
\end{equation}

$\mathbf{x}^{(k)}$ module corresponds to Eq. (\ref{eq:x}) and is used to calculate the $\mathbf{x}^{(k)}$ from $\mathbf{r}^{(k)}$ in the $k$-th iteration. A combination of two convolutional layers and a Rectified Linear Unit (ReLU, i.e. a diode) $\text{ReLU}(x)=\max(0, x)$ is employed to construct the sparse transformation $\mathcal{F}(\cdot )$ in Eq. (\ref{eq:x}), i.e., $\mathcal{F}(\mathbf{x})=\mathbf{B} \text{ReLU}(\mathbf{A} \mathbf{x})$, where $\mathbf{A}$ and $\mathbf{B}$ both use convolutional layers without bias terms to achieve equivalent matrix operations. To overcome the vanishing gradient problem due to the increased iteration blocks, which can lead to poorer performance of deep unfolding based CSI feedback,  a residual structure enhances recovery accuracy. 

Based on Eq.~(\ref{eq:x}), we assume that $\mathbf{x}^{(k)}=\mathbf{r}^{(k)}+\mathbf{w}^{(k)}+ \mathbf{e}^{(k)}$, where $\mathbf{w}^{(k)}$ represents the missing high-frequency components in $\mathbf{r}^{(k)}$, and $\mathbf{e}^{(k)}$ represents the noise. We then apply linear operation $\mathcal{R}(\cdot)$ to extract the missing component $\mathbf{w}^{(k)}$ from $\mathbf{x}^{(k)}$, i.e., $ \mathbf{w}^{(k)}=\mathcal{R}\left(\mathbf{x}^{(k)}\right)$. Define $\mathcal{R}(\cdot)$ as $\mathcal{R}=\mathcal{B}\circ\mathcal{M}$, where $\mathcal{M}$ and $\mathcal{B}$ corresponds to a convolutional layer without bias terms with
kernel size $3\times3$. 
It is known that when the sparse transformation satisfies $\mathcal{F}(\mathbf{x})=\mathbf{B} \text{ReLU}(\mathbf{A} \mathbf{x })$, the following approximation holds:
$\left\|\mathcal{F}(\mathbf{x})-\mathcal{F}\left(\mathbf{r}^{(k)}\right)\right\|^{2} \approx \alpha\left\|\mathbf{x}-\mathbf{r}^{(k)}\right\|^{2}$  \cite{istanet},
where $\alpha$ is a scalar only related to parameters of the sparse transformation $\mathcal{F}(\cdot)$.
Next, decompose $\mathcal{F}^{(k)}$ into $\mathcal{F}^{(k)}=\mathcal{H}^{(k)}\circ\mathcal{M}^ {(k)}$, where $\mathcal{H}^{(k)}$ consists of two convolutional layers without bias and a ReLU activation function. 
Eq.~(\ref{eq:x}) can be transformed into
% \begin{equation} \label{eq:xk2}
\begin{align} \label{eq:xk2}
    \mathbf{x}^{(k)}=\underset{\mathbf{x}}{\arg \min } &\frac{1}{2}\left\|\mathcal{H}^{(k)}(\mathcal{M}^{(k)}(\mathbf{x}))-\mathcal{H}^{(k)}\left(\mathcal{M}^{(k)}\left(\mathbf{r}^{(k)}\right)\right)\right\|^{2} \notag \\ 
    &+\theta^{(k)}\|\mathcal{H}^{(k)}(\mathcal{M}^{(k)}(\mathbf{x}))\|_{1}.
\end{align}
% \end{equation}
Next, construct the left inverse function of $\mathcal{H}^{(k)}(\cdot)$ such that $\widetilde{\mathcal{H}}^{(k)} \circ \mathcal{H}^ {(k)}=\mathcal{I}$, where $\mathcal{I}$ is the identity matrix operation. We then can use a DNN to construct a symmetric structure of $\widetilde{\mathcal{H}}^{(k)}(\cdot)$ as $\mathcal{H}^{(k)}(\cdot)$, and add the constraint of $\widetilde{\mathcal{H}}^{(k)} \circ \mathcal{H}^{(k)}=\mathcal{I}$ to the loss function. Finally, a closed-form 
expression of $\mathbf{x}^{(k)}$ can be obtained as
\begin{equation} \label{eq:xk3}
\mathbf{x}^{(k)}=\mathbf{r}^{(k)}+\mathcal{B}^{(k)}\left[\widetilde{\mathcal{H}}^{(k)}\left[
\mbox{soft}\left[\mathcal{H}^{(k)}\left(\mathcal{M}^{(k)}\left(\mathbf{r}^{(k)}\right)\right), \theta^{(k)}\right]\right]\right]
\end{equation}
where the soft threshold function is defined as $\mbox{soft}(x, \theta)=\operatorname{sgn}(x) \max (0, |x|-\theta)$.
The network structure corresponding to the $\mathbf{x}^{(k)}$ module is shown in the gray box at Fig. \ref{figure_architecture}, where kernel number $C$ is set to $32$ by default.

% \subsection{Initialization and Loss Function}
% When the gNB receives the compressed vector $\mathbf{y}$, it will reconstruct the pre-compressed CSI vector $\mathbf{x}$ through $N_I$ iteration modules. First, the initial estimate $\mathbf{x}^{(0)}$ is required to get $\mathbf{r}^{(1)}=g_{\Phi,\rho^{(1)}}\left(\mathbf{x}^{(0)}, \mathbf{y}\right)$. For $\mathbf{x}^{(0)}$ in the initial iteration, least squares estimation is utilized to the initialization, i.e., $\mathbf{x}^{(0)} = \mathbf{A}_{t}\mathbf{y}$, where $\mathbf{A}_{t} = \mathbf{XY}^{\top}\left(\mathbf{Y} \mathbf{Y} ^{\top}\right)^{-1}$ is the least square estimation of the transformation matrix $\mathbf{A}$ in the linear transformation $\mathbf{x} = \mathbf{A}\mathbf{y}$, $\mathbf{X}\in \mathbb{R}^{N\times N_T}$ is the matrix composed of all column vectors $\mathbf{x}$ in the training set, $\mathbf{Y}\in \mathbb {R}^{M\times N_T}$ is the matrix composed of all column vectors $\mathbf{y}$ in the training set, and $N_T$ is the number of samples in the training set.
% % Note that $\mathbf{A}_{t} = \mathbf{XY}^{\top}\left(\mathbf{Y} \mathbf{Y}^{\top}\right)^{-1}$ is A matrix that will use the same $\mathbf{A}_{t}$ as the training set for the test set to obtain the initial estimate $\mathbf{x}^{(0)}$. 
% Then the CSI matrix before compression can be recovered by iterating the neural network module corresponding to the Eq. (\ref{eq:rk}) and the Eq. (\ref{eq:xk3}).
% the neural network

To optimize the parameters of SPTM2-ISTANet+, an efficient loss function needs to be constructed for training. 
% The loss function of SPTM2-ISTANet+ consists of two parts, the MSE term used to optimize the CSI reconstruction accuracy and the one used to ensure $\widetilde{\mathcal{H}}^{(k)} \circ \mathcal{H} ^{(k)}=\mathcal{I}$ constraints. 
Define the size of the training data set as $N_T$, the $i$ sample in the training set as $\mathbf{x}_i \in \mathbb{R}^{N}$, and the number of iteration modules as $N_I$. The loss function can be constructed as
\begin{equation}
\mathcal{L}_{\text {total}}(\Theta)=\mathcal{L}_{\text {MSE}}+\gamma\cdot \mathcal{L}_{\text {constraint}},
\end{equation}
where $\mathcal{L}_{\text {MSE}}=\frac{1}{N_{T} N} \sum_{i=1}^{N_{T}}\left\|\mathbf{ x}_{i}^{\left(N_{I}\right)}-\mathbf{x}_{i}\right\|^{2}$ is the CSI reconstruction accuracy indicator - mean square error (MSE), which is commonly used in the CSI feedback,
$\mathcal{L}_{\text {constraint}} = \frac{1}{N_{T}N} $ $ \sum_{i=1}^{N_{T}} \sum_{k= 1}^{N_{I}}\left\|\widetilde{\mathcal{H}}^{(k)}\left(\mathcal{H}^{(k)}\left(\mathcal{M} ^{(k)}\left(\mathbf{r}_i^{(k)}\right)\right)\right) - \mathcal{M}^{(k)} \\ \left(\mathbf{r} _i^{(k)}\right)\right\|^{2}$ corresponds to $\widetilde{\mathcal{H}}^{(k)} \circ \mathcal{H}^{(k)}= \mathcal{I}$ restriction, and $\gamma$ is the regularization weight, which is set to $0.01$ here.

\section{Model-Aided Data Augmentation}
\vspace*{-2mm}
\begin{figure}[thpb]
      \centering
      \includegraphics[scale=0.26]{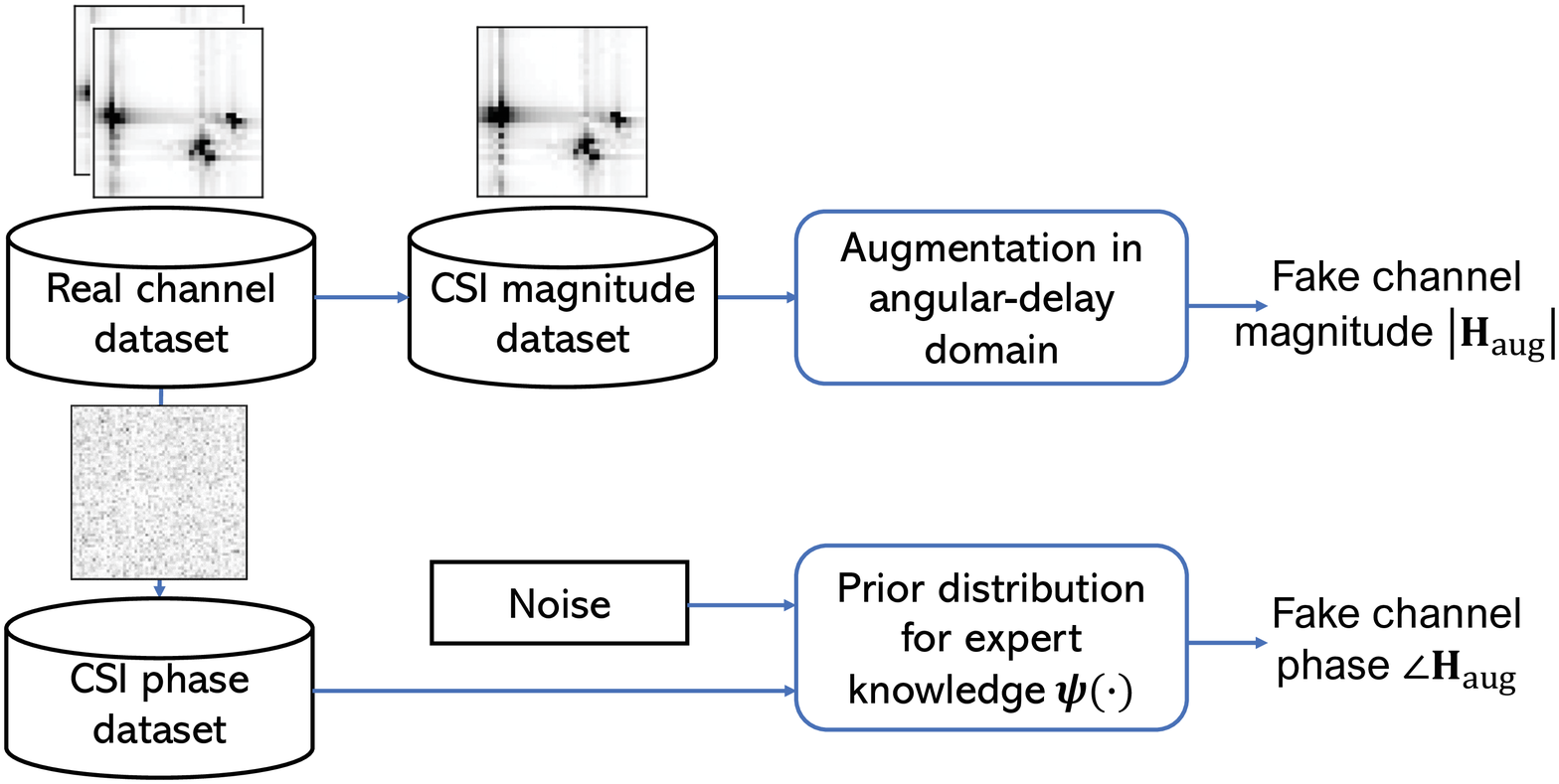}\vspace*{-1mm}
    %   \caption{Architecture comparison between proposed data augmentation models and GAN-base data augmentation.}
      \caption{Proposed CSI data augmentation strategy.}
      \label{figure_gan_org}
    %   \vspace*{-2mm}
  \end{figure}
  
% As shown in Fig. \ref{figure_gan_org}(a), traditional GAN alternately optimizes the generator and the
% discriminator in competition with each other by iteratively
% and alternately solving the max and min optimization problems during training phase, i.e., minimizing the distance between the distributions of
%  fake channel and real channel. For the generator, a deep neural network
% is constructed to map the noise vector sampled with gaussian
% distribution to the generated fake channel. When the GAN arrives convergence with matched real channel measurement and fake channel distribution, the generator can be regarded as a stable
% storage of channel model and used to generate the fake channel
% data to form a extensive training dataset to support CSI feedback.

% GAN-based channel data augmentation alternately solves the maximizing and minimizing optimization problems during training to reduce the distance between the distributions of fake channel and real channel.
% However, it is still hard to converge the training of GAN using limited samples. Besides, when the available channel samples only covers the partial channel features in the required areas, the well-trained generator will be a highly biased channel model which can decrease the CSI feedback performance heavily. 
% Note that, precise channel distribution in the limited measurement samples is not enough, and the generated augmented samples should cover the features which is not available in the existing measurements.

To enhance DNN training for CSI feedback, we note that augmented samples should present unavailable or under-represented features among existing measurements, and develop a simple but effective model-driven augmentation by decoupling the characteristics in the magnitude and phase of CSI matrices. 
%  considering the above challenges, different from the previous channel augmentation works, which are not available also

To begin, we first split the phase and magnitude of MIMO channel matrices for channel augmentation, i.e., 
\begin{equation}
\mathbf{H}=|\mathbf{H}| \odot e^{j \angle \mathbf{H}},
\end{equation}
where $\odot$ represents Hadamard product, the $(m, n)$-th entry of $\mathbf{H}$ is denoted as $\mathbf{H}_{m, n}=\left|\mathbf{H}_{m, n}\right| e^{j \angle \mathbf{H}_{m, n}}$, the magnitude matrix is denoted by $|\mathbf{H}|$ with entries $\left|\mathbf{H}_{m, n}\right|$ and phase matrix is denoted as $\angle \mathbf{H}$ with entries $\angle \mathbf{H}_{m, n}$. 
In this way, prior knowledge in existing works about CSI features in the multipath profile and phase distribution can be exploited easily.

Next, we utilize the geographical continuity of CSI variation to generate augmented magnitude matrices, which should exhibit similar characteristics to the measured channels. Given a static environment and fixed paths between the gNB and the UE, it has been shown that any geographical continuity of UE movement should lead to continuous variation in the angular-delay domain \cite{dyloc}. In other words, CSIs in the vicinity of a measured point are highly correlated in the angular-delay domain considering the influence of angle of arrival/departure and multipath delays. Consequently, we can construct consecutive and multiple angle-delay profiles by shifting the CSI magnitude matrix in the angular-delay domain to form a space-series data, which can reflect the features of nearby UE channels.
We further utilize the circular characteristic of CSI matrices in the angular-delay domain owing to the property of DFT, and make the circular shiftings. The entries of the augmented matrix $\left|\mathbf{H}^\text{aug}_{m, n}\right|$  are characterized by the following rule in the angular-delay domain, i.e., $\forall m\in\left[R_{d}\right], n \in\left[N_{b}\right]$, 
% \begin{equation}
% \left|\mathbf{H}^\text{aug}_{m, n}\right|=\left|\mathbf{H}_{m, (n+i)\bmod N_b}\right|,\forall m \in\left[R_{d}\right], n \in\left[N_{b}\right],
% \end{equation}
% and the rule in the delay domain owing to the trucation:
\begin{equation}
    \left|\mathbf{H}^\text{aug}_{m, n}\right|=\left\{\begin{matrix}\left|\mathbf{H}_{m+i,(n+j)\bmod N_b}\right|,
 1\le m+i\le R_d,\\
0, \text{else}, %\text{if }
\end{matrix}\right. 
\end{equation}
where $\lfloor -\frac{R_{d}}{2}\rfloor \le i \le  \lfloor \frac{R_{d}}{2}\rfloor$ and $\lfloor -\frac{N_{b}}{2}\rfloor \le j \le  \lfloor \frac{N_{b}}{2}\rfloor$ are the shifting step length in angular and delay domain, respectively. We apply truncation in the delay domain by setting to
zero delay elements beyond $R_d$ rows.

Next, phase matrices are augmented. Assuming the phase distribution from the limited CSI measurements is $\boldsymbol{\psi}_\text{measure}(\cdot)$.  Existing works show that the phase of the channel coefficient appears to be uniformly distributed over $[0, 2\pi]$ for the narrow-band MIMO channels \cite{yu2004models, jw2003models}. Additionally, the augmented phases can cover cases beyond the measured CSIs to enhance the training and avoid overfitting. We select the uniform distribution as the augmented phase distribution $\boldsymbol{\psi}_\text{aug}(\cdot) \sim \mathcal{U}(0, 2\pi)$. In other words, we construct a larger phase constraint space than the limited measurements, and use recall to replace some precision to enhance 
recovered CSI accuracy in the real deployment.
Accordingly, we replace each phase element in a sampled CSI matrix with a random phase, i.e.,
\begin{equation}
 \angle\mathbf{H}^{\text {aug}}_{m, n}= \angle e^{-j \theta_{m,n}} , \forall m \in\left[R_{d}\right], n \in\left[N_{b}\right],
\end{equation}%\mathbf{H}^{(m, n)}
where $\theta_{m,n} \sim \mathcal{U}(0, 2\pi)$.
Finally, we combine augmented magnitude matrices and phase matrices to generate the full CSI augments.

% We further use the Kolmogorov–Smirnov test \cite{ref:kstest} to check the fitness of uniform distribution and phase distribution from the channel model. We use the Cost2100 channel simulator and 3GPP 38.901 UMa from the Quadriga simulator to generate two datasets which all have $10,000$ samples. For the distribution of $\angle\mathbf{H}^{(m, n)}, \forall m \in\left[R_{d}\right], n \in\left[N_{b}\right]$, they all accept the uniform hypothesis at the 5\% significance level. For the distribution with the matrix $\angle\mathbf{H}$, the Cost2100 dataset accepts $10.9\%$ uniform hypothesis at the 5\% significance level, and the Quadirga dataset accepts $8.3\%$ uniform hypothesis at the 5\% significance level. Consequently, uniform distribution can be a good way for the phase augmentation.

% \begin{figure}[thpb]
%       \centering
%       \includegraphics[scale=0.23]{hybridgan.pdf}\vspace*{-1mm}
%       \caption{Architecture of proposed data augmentation models.}
%       \label{figure_gan}
%       \vspace*{-3mm}
%   \end{figure}

\vspace*{-1mm}
\section{Performance Evaluation}

\subsection{Experiment Setup}
\vspace*{-1mm}
% For both training and testing, 
% we use CSI matrices generated by the COST 2100 model \cite{c2100} and Quagirga. 
% Two scenarios are tested: (a) outdoor channel
% with 300~MHz downlink bands using Cost2100; 
% (b) outdoor channel
% with 2.1~GHz downlink bands using 3GPP 38.901 UMa from the Quadriga simulator. 
% gNBs are placed at the center of a square area of length $400$m. Both  downlink bandwidths are $20$ MHz. 
% We randomly position UEs within the coverage area. The
% gNB uses ULA with $N_b = 32$ antennas and $N_f = 1024$ subcarriers. 
% After transforming the channel matrix into the delay domain, 
% only the first 32 rows are kept due to sparsity. 
% The overall training sample size is $100,000$ and testing sample size is $20,000$. For the data augmentation case, the samples given the limitet measurement number will be select from the training set.
% The values of epoch and batch size are set to $200$ and $64$, respectively.

For performance evaluation, 
we view CSI matrices generated by the COST2100 model \cite{c2100} as measurement data 
Since DL-based CSI feedback works have already achieved satisfactory performance in the indoor scenario \cite{dl_multires, guo2019convolutional,tang2021dilated, ref:SRNet}, we focus on the outdoor scenario where
practical CSI measurement is harder and CSI recovery performance is less satisfactory.
We test the following outdoor scenario commonly used: outdoor channel using a 300~MHz downlink, served by a gNB at the center of a square area of length $400$m with the bandwidth of $20$ MHz. 
We give $N_b=32$ antennas and $N_f = 1024$ subcarriers to the gNB to
serve single antenna UEs randomly distributed within
the coverage area. 
After transforming the CSI matrix into the angular-delay domain, 
only the first 32 rows are kept owing to sparsity. 
The overall training set size is $100,000$ and the testing set size is $20,000$. For data augmentation, samples from the limited measurements will be randomly selected from the training set. The shifting ranges in the angular domain and delay domain are $-15$ to $15$ and $-3$ to $3$, respectively. The training dataset size after augmentation is set to equal the overall training set size by repetition or phase randomization when the initial augmentation is not large enough (depending on the corresponding augmentation method). We use 200 epochs and a batch size of $64$.

To compare recovery accuracy of different networks, we adopt the metric
$\textrm{NMSE} = \frac{1}{n}
\sum_{k=1}^n\Arrowvert\mathbf{H}_k-\mathbf{\hat{H}}_k\Arrowvert^2/\Arrowvert\mathbf{H}_k\Arrowvert^2,
$
where $\mathbf{\hat{H}}$ is the recovered $\mathbf{H}$, $k$ and $n$ are the index 
and number of samples in the testing set, respectively.

% Compressed sensing algorithms such as LASSO \cite{ref:lasso-l1}, BM3D-AMP \cite{ref:bm3d-amp} and TVAL3 \cite{ref:tval3} have been verified in \cite{ref:csinet} that the performance is worse than the earlier version of CsiNet+ \cite{guo2019convolutional}, so the performance of related schemes will not be compared here. 

\subsection{Feedback Accuracy Comparison}
\vspace*{-1mm}
We compare SPTM2-ISTANet+ with two schemes of superior performance in massive MIMO system CSI feedback (without relying on additional auxiliary information, e.g., uplink CSI, CSI at previous moments):
\begin{itemize}
    \item CsiNet+ \cite{guo2019convolutional}, which enhances the CSI feedback performance by using larger convolution kernels $(7\times7)$ and optimizing the structure of residual units.
    \item DCRNet \cite{tang2021dilated}, which is an enhanced design of CRNet \cite{dl_multires} and combines multiple resolution convolution kernels and dilated convolutions to extract the different-granularity features of the CSI matrix, and optimizes learning rate adjustment with the help of a warm-up process.
\end{itemize}
 We also use the vanilla ISTANet+ \cite{istanet} designed for image processing as a benchmark, where the orthogonal random Gaussian measurement matrix is adopted as $\Phi \in \mathbb{R}^{M \times N}$, where $N = 2048$, $M$ is determined by CR. 
TM2-ISTANet+ is the SPTM2-ISTANet+ without spherical processing to show the advantage of different modifications in SPTM2-ISTANet+.
% For SPTM2-ISTANet+, $\Phi \in \mathbb{R}^{(M-1) \times 2048}$ owing to the influence of spherical processing. 

\begin{figure} \centering 
\includegraphics[width=0.96\columnwidth]{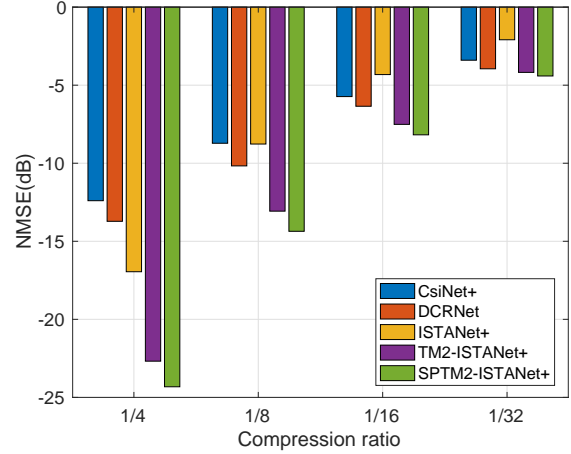} 
\caption{NMSE comparison in different CRs.} 
\label{figure_nmse} 
\vspace*{-5mm}
\end{figure}

\begin{figure} \centering 
\includegraphics[width=0.96\columnwidth]{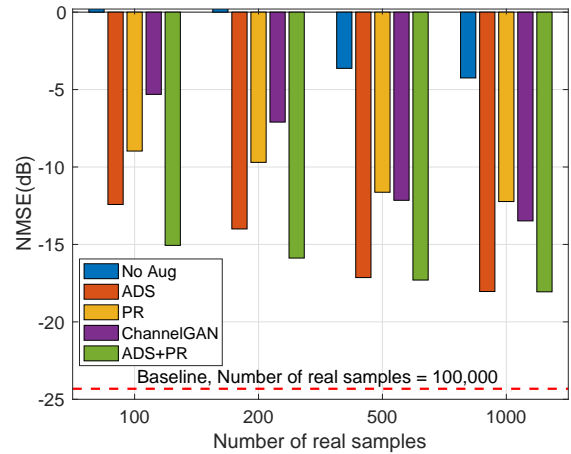} 
\caption{NMSE comparison using different data augmentation strategies when CR = $\frac{1}{4}$.} 
\label{fig:figure_aug} \vspace*{-4mm}
\end{figure}

Fig. \ref{figure_nmse} shows the CSI feedback performance comparison among the five schemes CsiNet+, DCRNet, ISTANet+, TM2-ISTANet+ and SPTM2-ISTANet+ at different CRs for outdoor scenario. The number of iteration blocks is set to 9. As shown in Fig. \ref{figure_nmse}, our proposed SPTM2-ISTANet+ can achieve better performance than CsiNet+, DCRNet and ISTANet+ at all tested CRs. Especially when CR is $\frac{1}{4}$ in the outdoor scenario, SPTM2-ISTANet+ can achieve about $10$dB improvement in CSI reconstruction accuracy compared to the traditional deep learning schemes DCRNet and CsiNet+ (i.e., NMSE decreases by $10$dB). On the other hand, we notice that ISTANet+ performs better than CsiNet+ and DCRNet for larger
CR (such as $\frac{1}{4}$).
The performance of ISTANet+ is worse than that of CsiNet+ and DCRNet at higher compression 
(or lower CR), e.g., $\frac{1}{32}$. Results show that the random measurement matrix $\Phi$ used in traditional CS methods does not effectively extract key data features when CR is small, while TM2-ISTANet+ based on the data-driven measurement matrix does better. Moreover, we also notice the benefits of SPTM2-ISTANet+ 
attributed to the spherical processing in all CRs. 
% Finally, it is worth mentioning that compared with general CSI feedback networks including CsiNet+ and CRNet, SPTM2-ISTANet+ does not require the use of batch normalization, sigmoid or tanh activation functions, which have poor interpretability. 

\subsection{Augmentation Performance Comparison}
\vspace*{-1mm}

Fig. \ref{fig:figure_aug} shows the performance of SPTM2-ISTANet+ using different augmentation strategies including the costly ChannelGAN \cite{ref:GAN_xiao}, no augmentation (No Aug) which uses repetition to enlarge dataset size, shifting in angular-delay domain (ADS) where repetition is used to enlarge dataset size if necessary, phase randomization (PR), and ADS together with RP (ADS+PR). We set the CR to $\frac{1}{4}$. 
We select the number of CSI measurements before augmentation to $100$, $200$, $500$, and $1000$, respectively. As shown in Fig. \ref{fig:figure_aug}, our proposed ADS and ADS+PR can significantly outperform ChannelGAN in each case, where PR alone can outperform ChannelGAN when the number of CSI measurement samples is below 500. Notably, ADS+PR can achieve NMSE of $-15$dB using only $100$ CSI measurement samples whereas ChannelGAN mainly achieves $-5.3$dB. Furthermore, both ADS and PR can help improve the CSI recovery accuracy.
Overall, the proposed low-cost training enhancement by using ADS always achieves higher gains than PR owing to the better utilization of geographical correlation.

\begin{table}[]
\renewcommand{\arraystretch}{1.5}
\centering
\caption{FLOPs of encoding networks in UE. M: million.}
\label{tab:comp-complex}
\begin{tabular}{|c|c|c|c|}
\hline
% & \multicolumn{1}{c|}{\textbf{CsiNet-LSTM}} & \multicolumn{1}{c|}{\textbf{MarkovNet}} & \multicolumn{1}{c|}{\textbf{MarkovNet-CNN}} \\ \hline
  & \textbf{CsiNet+} & \textbf{DCRNet} & \textbf{SPTM2-ISTANet+} \\ \hline
\textbf{CR=$\frac {1}{4}$}  &  2.9 M            & 2.6 M         & 2.1 M         \\ \hline
\textbf{CR=$\frac {1}{8}$} &  1.9 M            & 1.6 M          & 1 M         \\ \hline
\textbf{CR=$\frac {1}{16}$} &  1.3 M            & 1 M         & 0.5 M         \\ \hline
\textbf{CR=$\frac {1}{32}$} &  1.1 M           & 0.8 M         & 0.3 M         \\ \hline
\end{tabular}\vspace*{-3mm}
\end{table}

\begin{table}[!hbtp]
\renewcommand{\arraystretch}{1.5}
\caption{Parameters and computational complexity of augmentation strategies. B: Billion, M: Million, K: thousand.}
\begin{center}
\begin{tabular}{|c|c|c|c|c|}
\hline
\textbf{} & \textbf{ChannelGAN}  & \textbf{ADS} & \textbf{PR} & \textbf{ADS+PR} \\ \hline
\textbf{Parameters}     &  11.7 M   & 0.2 K & 1 K & 1 K  \\ \hline
\textbf{FLOPs}      &   5.4 B  &  -  & 4.1 K & 4.1 K  \\ \hline
\end{tabular}\vspace*{-5mm}
\end{center}
\label{tab:aug-comp-complex}
\end{table}

% We demonstrate that latent convolutional layers 
% require
% significantly fewer parameters than FC-layers 
% without loss of performance. Table~\ref{tab:comp-complex}
% compares the model size and computational 
% complexity (respectively) of CsiNet-LSTM, MarkovNet, and MarkovNet-CNN associated with a single timeslot. 
% Among the tested compression ratios,
% MarkovNet uses $\frac{1}{60}$ of the parameters 
% in comparison to CsiNet-LSTM. 
% More importantly, MarkovNet-CNN further reduces 
% the number of parameters to $\frac{1}{3000}$ of what is
% needed by CsiNet-LSTM 
% while achieving similar or
% better CSI recovery accuracy. 
% We further provide the parameters of 
% several related networks that do not exploit temporal correlation (CsiNet \cite{ref:csinet}, CRNet\cite{ref:Lu2020CRNet}, and Deep AE\cite{ref:Jang2019}) 
% in Table~\ref{tab:comp-complex} for a more comprehensive comparison. We observe 
% MarkovNet uses similar number of parameters, whereas MarkovNet-CNN 
% requires significantly fewer parameters with the help of the proposed CNN-based dimension compression and decompression modules.
\addtolength{\topmargin}{0.03in}
\subsection{Complexity Comparison}
\vspace*{-1mm}

Table~\ref{tab:comp-complex} compares FLOPs of encoding networks in UE, the decoding networks are neglected since the computing power and energy of the resource-rich gNB are generally of less concern. SPTM2-ISTANet+ can reduce UE computation by over $27\%$ and $19\%$ in comparison with CsiNet+ and DCRNet when CR is $\frac{1}{4}$, respectively, and more computations can be saved as CR decreases. Parameters of the above three encoding networks are at a similar level, and can refer to \cite{guo2019convolutional}.

Table~\ref{tab:aug-comp-complex} compares parameters and FLOPs of different augmentation strategies. Unlike ChannelGAN which requires millions of parameters and billions of FLOPs, our proposed ADP+RP only needs thousands of parameters and FLOPs to achieve a higher CSI recovery accuracy.

% \subsection{Encoder Complexity Comparison}

% \vspace*{-2mm}
\section{Conclusions}
In this paper,  we develop a training enhancement solution for DL-based massive MIMO CSI feedback with small datasets to improve the CSI feedback efficiency and accuracy.  We propose a simple and effective data augmentation strategy by augmenting the limited channel measurements based on domain knowledge and physical insight into wireless channel characteristics.
We develop an efficient deep unfolding-based CSI feedback network SPTM2-ISTANet+, particularly for the challenging outdoor environment.  By decoupling the features in the magnitude and phase of CSI matrices, our proposed augmentation strategy together with SPTM2-ISTANet+ can significantly enhance CSI recovery performance, and can achieve NMSE of $-15$dB by using only $100$ measurement channel samples when CR is $\frac{1}{4}$.

\addtolength{\textheight}{-5cm}   % This command serves to balance the column lengths

\ifCLASSOPTIONcaptionsoff
  \newpage
\fi

\bibliographystyle{IEEEtran}
\bibliography{ref}

\end{document}